\documentclass[aps,prl,twocolumn,superscriptaddress,showpacs,floatfix]{revtex4-1}
\usepackage{graphicx,bm,epsfig,textcomp,color,dcolumn,setspace,array,mathrsfs,amsmath,amssymb,gensymb,booktabs,url,hyperref,bibentry,natbib}
\begin{document}

\title{Magnonic Su-Schrieffer-Heeger Model in Honeycomb Ferromagnets}

\author{Yu-Hang Li}
\affiliation{Department of Electrical and Computer Engineering, University of California, Riverside, California 92521, USA}
\author{Ran Cheng}
\email[]{ran.cheng@ucr.edu}
\affiliation{Department of Electrical and Computer Engineering, University of California, Riverside, California 92521, USA}
\affiliation{Department of Physics and Astronomy, University of California, Riverside, California 92521, USA}

\begin{abstract}
Topological electronics has extended its richness to non-electronic systems where phonons and magnons can play the role of electrons. In particular, topological phases of magnons can be enabled by the Dzyaloshinskii-Moriya interaction (DMI) which acts as an effective spin-orbit coupling. We show that besides DMI, an alternating arrangement of Heisenberg exchange interactions critically determines the magnon band topology, realizing a magnonic analog of the Su-Schrieffer-Heeger model. On a honeycomb ferromagnet with perpendicular anisotropy, we calculate the topological phase diagram, the chiral edge states, and the associated magnon Hall effect by allowing the relative strength of exchange interactions on different links to be tunable. Including weak phonon-magnon hybridization does not change the result. Candidate materials are discussed.

\end{abstract}
\maketitle

\textit{Introduction.---}
Magnonics, which utilizes the spin-wave excitations (\textit{i.e.} magnons) in magnetic insulators rather than the spin of electrons to transfer spin angular momenta, has attracted persistent attentions in physics and engineering~\cite{Chumak2015Magnon,Demokritov2012Magnonics,Nikitov2015Magnonics}. Different from electrons, magnons are charge neutral so that generating a magnon spin current does not incur Joule heating, which holds huge potential to realize low-dissipation devices~\cite{Ruckriegel2018Bulk,Awad2017Long,Cornelissen2015Long}.

In the past decade, an emerging direction in magnonics has been the study of topological properties of magnon bands and the associated exotic transport phenomena. In a ferromagnetic Kagome lattice, the Dzyaloshinskii-Moriya interaction (DMI) plays the role of an effective spin-orbit coupling (SOC) and opens a topological nontrivial gap in the magnon spectrum~\cite{Katsura2010Theoty}. Consequently, a longitudinal temperature gradient can induce a transverse magnon current due to the Berry phase effect, leading to the magnonic counterpart of the quantum anomalous Hall effect~\cite{Onose2010Observation,Matsumoto2011Theoretical,Matsumoto2011Rotational,Xiao2006Berry,Zhang2013Topological}.

The discovery of two dimensional magnets brings about new exciting opportunities to explore topological magnons~\cite{Huang2017Layer,Lee2016Ising,Wang2016Raman,Gong2017Discovery}. For instance, a honeycomb ferromagnet with perpendicular order displays Dirac-type magnon dispersions around the $K$ (and $K^{\prime}$) point~\cite{Yelon1971Renormalization,amuelsen1971Spin,Pershoguba2018Dirac}. Meanwhile, its special crystal symmetry allows for the second-nearest neighboring DMI, which acts as a Rashaba-type SOC on magnons, resulting in a magnonic analog of topological insulators~\cite{Chen2018Topological,Kim2016Realization,Owerre2016A,Owerre2016Topological}. In addition, topological effect manifests in honeycomb antiferromagnets as the spin Nernst effect, where an in-plane temperature gradient generates a pure spin current in the transverse direction without a magnon Hall current~\cite{Cheng2016Spin,Zyuzin2016Magnon,Shiomi2017Experimental}.

\begin{figure}[t]
  \centering
  \includegraphics[width=0.9\linewidth]{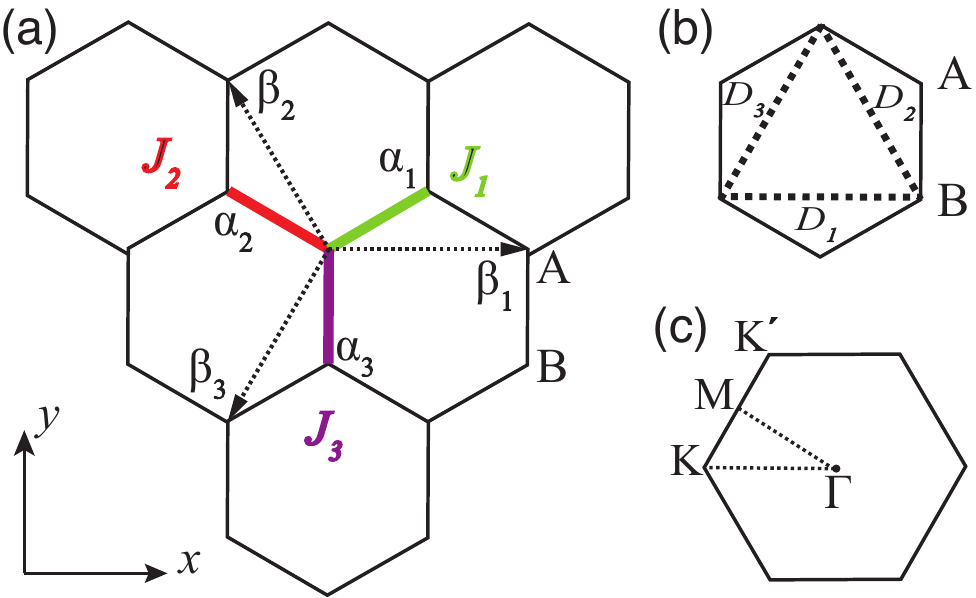}
  \caption{(color online).
  	(a) Schematics for the magnonic SSH model in a honeycomb ferromagnet. The NN and second-NN bonds are labeled by $\bm{\alpha}_{i}$ and $\bm{\beta}_{i}$, respectively. The NN exchange interaction along $\bm{\alpha}_{i}$ direction is $J_{i}$. 
	(b) The second-NN DMI on different bonds.
	(c) The first Brillouin zone of the reciprocal lattice. 
  	}
\label{schematic}
\end{figure}

In spite of remarkable progress in topological magnons, phase transitions among magnonic states of distinct topology have so far been restricted to the variation of DMI~\cite{Kim2016Realization,Owerre2016Topological}, temperature\cite{Kim2016Realization}, and magnon-phonon coupling~\cite{Zhang2020SU,Thingstad2019Chiral}. An important missing ingredient is the anisotropy among nearest-neighboring (NN) exchange interactions on different atomic bonds, as illustrated in Fig.~\ref{schematic}(a). While the DMI amounts to a magnonic SOC, the NN exchange interactions play the role of hopping parameters for magnons. The well-known Su-Schrieffer-Heeger (SSH) Model of electrons reveals a profound relation between an alternating hopping amplitudes and the band topology in one dimensions~\cite{Su1980Soliton}. When generalized into a honeycomb lattice, there can be three different NN hopping parameters forming an alternating pattern, giving rise to intriguing topological phases in two dimensions. At this point, it is tempting to ask if a magnonic analog of the two-dimensional SSH model exists and, more importantly, what non-trivial transport phenomena are enabled in such a system.

In this Letter, we study a honeycomb ferromagnet with tunable NN exchange interactions $J_1$, $J_2$, and $J_3$ as illustrated in Fig.~\ref{schematic}(a). By varying the relative strength of $J_1$, $J_2$, and $J_3$, we demonstrate that the system can undergo phase transitions between a magnon Hall insulator (MHI) with a pair of chiral edge modes and a trivial magnon insulator without any edge modes. This topological phase transition is characterized by a sharp change of the magnon Hall coefficient, which can be detected experimentally. We also consider weak hybridization of magnons and phonons, which, though inevitable in thermal transport, does not change our essential conclusion. Finally, our proposed topological phase transition can be tested in monolayer transition metal trichalcogenides subject to lattice deformation (\textit{e.g.}, by straining).

\textit{Model.---}
Let us consider a collinear ferromagnet on a honeycomb lattice as schematically shown in Fig.~\ref{schematic} (a). The minimal Hamiltonian of such a system is
\begin{align}
\mathcal{H}=-\sum_{\left<i,j\right>}J_{ij}\bm{S}_{i}\cdot\bm{S}_{j}+D\sum_{\left<\left<i,j\right>\right>}\epsilon_{ij}\hat{\bm{z}}\cdot\bm{S}_{i}\times\bm{S}_{j}-\kappa\sum_{i}S_{iz}^{2},
\label{model1}
\end{align}
where $J_{ij}>0$ are the variable NN exchange interactions, $\kappa>0$ is the easy-axis anisotropy, and $D$ is the second-NN DMI with $\epsilon_{ij}=\pm 1$ depending on the chirality of atomic links~\cite{Owerre2016Topological,Owerre2016A,Kim2016Realization,Cheng2016Spin}. Specifically, the NN exchange interactions include $J_1$ along ${{\alpha}}_1$,  $J_2$ along ${{\alpha}}_2$, and $J_3$ along ${{\alpha}}_3$ directions, respectively. The difference among $J$ can be realized by lattice straining, which is usually controllable via a voltage. With lattice deformation, the second-NN DMI can also exhibit directional anisotropy as illustrated in Fig.~\ref{schematic}(b). However, as discussed in the supplementary materials (SI)~\cite{Supply}, the anisotropy of DMI does not affect the band topology, introducing only trivial modifications to the band dispersion. For simplicity, we will treat the DMI as isotropic throughout this Letter.

Using the Holstein-Primakoff transformation~\cite{Nolting2009Quantum}, we set $S_{iA}^{+}=\sqrt{2S}a_i$, $S_{iA}^{-}=\sqrt{2S}a_i^{\dagger}$, $S_{iA}^{z}=S-a_i^{\dagger}a_i$ for the A sublattice, and $S_{iB}^{+}=\sqrt{2S}b_i$, $S_{iB}^{-}=\sqrt{2S}b_i^{\dagger}$, $S_{iB}^{z}=S-b_i^{\dagger}b_i$ for the B sublattice, where $a_i^\dagger$ ($b_i^\dagger$) creates a magnon on the A (B) sublattice in the $i$ unit cell. Under the basis $\psi\left(\bm{k}\right)=\begin{bmatrix}a_{\bm{k}},&b_{\bm{k}}\end{bmatrix}^T$ with $a_{\bm{k}}$ ($b_{\bm{k}}$) the Fourier transformation of $a_i$ ($b_i$) and $\bm{k}$ belonging to the first Brillouin zone depicted in Fig.~\ref{schematic}(c), Eq.~\eqref{model1} becomes $\mathcal{H}=\sum_{\bm{k}}\psi\left(\bm{k}\right)^{\dagger}H\left(\bm{k}\right)\psi\left(\bm{k}\right)$, where, after discarding the zero-point energy, the matrix $H(\bm{k})$ reads
\begin{align}
H\left(\bm{k}\right)=a\mathrm{I}+b\sigma_{x}+c\sigma_{y}+d\sigma_{z}.
\label{model2}
\end{align}
Here $a=(J_1+J_2+J_3+K)S$ with $K=\kappa\left(2S-1\right)/S$, $b+{\text{i}}c=-S\sum_i J_{i}e^{i\bm{k}\bm{\alpha}_{i}}$, $d=SD\sum_i{{\text{sin}}\left(\bm{k}\beta_i\right)}$ where $i=1,2,3$ and the linking vectors $\bm{\alpha}_i$, $\bm{\beta}_i$ are illustrated in Fig.~\ref{schematic}(a). In Eq.~\eqref{model2}, $\sigma_{x,y,z}$ are Pauli matrices acting on the sublattice space and $\mathrm{I}$ is the $2\times2$ unit matrix. The band topology does not depend on $a$ in Eq.~\eqref{model2}, which only causes an overall shift to the entire band structure. When $J_{1}=J_{2}=J_{3}$, the system becomes a bosonic Haldane model if $D$ is nonzero~\cite{Kim2016Realization,Owerre2016A,Owerre2016Topological}.

To solve the eigenvalue problem, we parameterize $b=l\sin{\theta} \cos{\psi}$, $c=l\sin{\theta}\sin{\psi}$, and $d=l\cos{\theta}$ with $l=\sqrt{b^2+c^2+d^2}$. Then diagonalizing Eq.~\eqref{model2} gives the eigenvalues $\hbar\omega_{\pm}=a\pm l$ and the eigenvectors $\psi_{+}=\begin{bmatrix}-\sin{\frac{\theta}{2}},&\cos{\frac{\theta}{2}} e^{-i\psi}\end{bmatrix}^{T}/\sqrt{2}$ and $\psi_{-}=\begin{bmatrix}\cos{\frac{\theta}{2}},&\sin{\frac{\theta}{2}} e^{-i\psi}\end{bmatrix}^{T}/\sqrt{2}$. The two magnon bands $\hbar\omega_{\pm}$ are mirror symmetric about the energy plane $\hbar\omega=a$. As magnons are bosonic excitations, the lower band $\hbar\omega_-$ is always more populated than the upper band $\hbar\omega_+$. We will therefore focus on the lower band when querying on the band topology even though both bands will be considered when calculating the transport coefficient. To simplify our discussion, we adopt the scaling convention that $J_1=1$, the hexagon side length $a_{0}=1$ such that other quantities are expressed dimensionlessly with respect to these parameters. In addition, we scale temperature $T$ by the Curie temperature $T_c=J_1S(S+1)/k_B$ evaluated by the mean-field theory at $J_1=J_2=J_3$. In the following, unless otherwise specified, $S=5/2$, $D=0.1$ and $\kappa=0.05$  whereas $J_2$ and $J_3$ are tunable.

\begin{figure}[t]
  \centering
  \includegraphics[width=\linewidth]{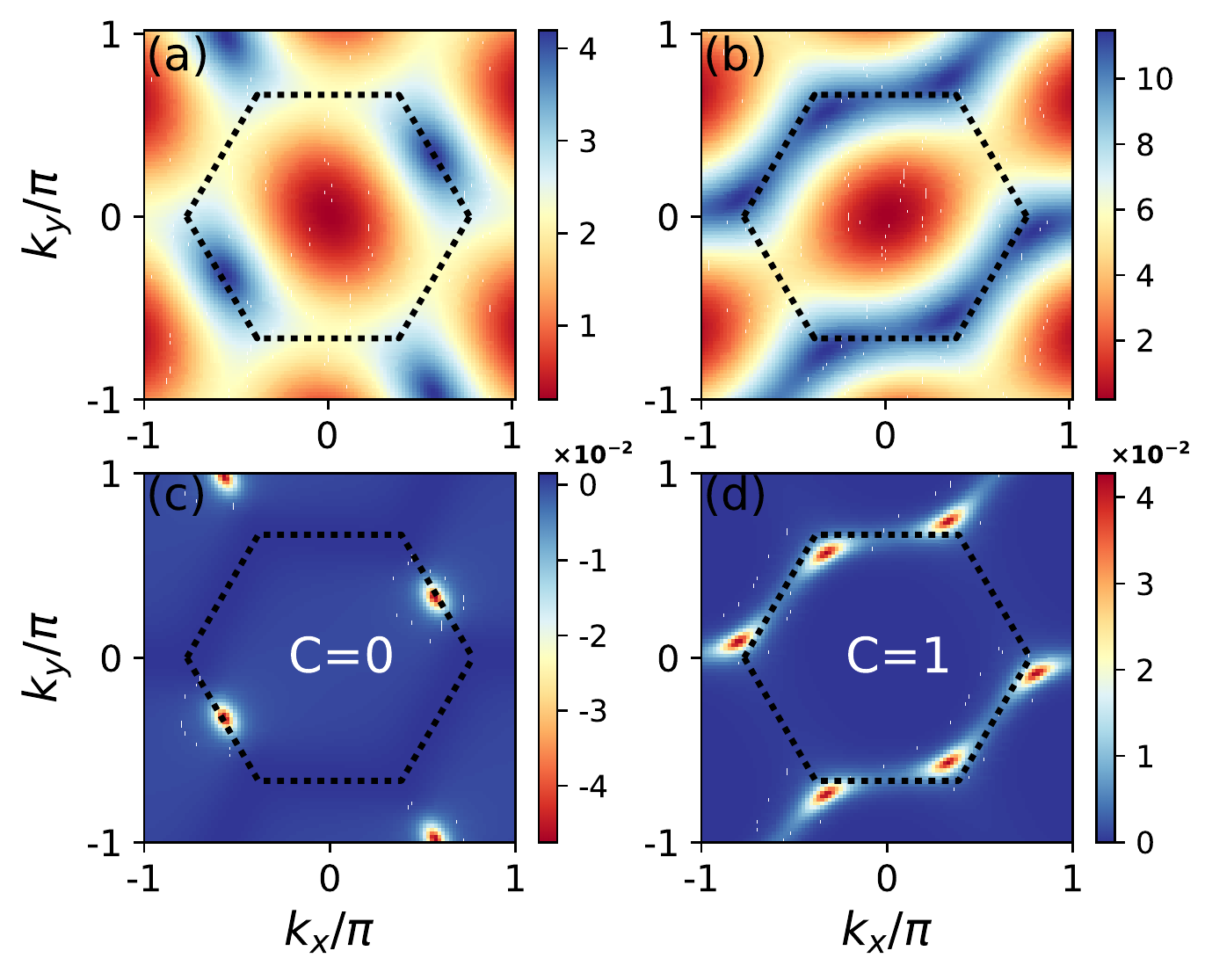}
  \caption{(color online).
  	(a) and (b): the dispersions of the lower magnon band for $J_2=J_3=0.4$ and $J_2=J_3=2$, respectively.
	(c) and (d): corresponding Berry curvatures, whose Chern numbers are $C=0$ and $C=1$.
	The dotted black lines denote the edges of the first Brillouin zone.
  	}
\label{bc}
\end{figure}

\textit{Band topology.---}
Key to our proposal is identifying different topological phases by varying the NN exchange interactions. It is thus instructive to first consider two representative cases. Figs.~\ref{bc}(a) and (b) plot the lower magnon bands $\hbar\omega_-$ for $J_2=J_3=0.4$ and $J_2=J_3=2$, respectively. The upper band, as mentioned previously, can be obtained directly by a mirror reflection about the $\hbar\omega=a$ plane. In both cases, the band structure exhibits a $C_2$ rotational symmetry ascribing to the anisotropy of the NN exchange interactions. The corresponding Berry curvature is $\bm{\Omega}_{\pm}\left(\bm{k}\right)=-\text{Im}\left<\nabla\psi_{\pm}\left(\bm{k}\right)\right|\times\left|\nabla\psi_{\pm}\left(\bm{k}\right)\right>=\mp\frac12\sin{\theta}\left(\nabla\theta\times\nabla\psi\right)$, which points to the $z$ direction. In Fig.~\ref{bc} (c) and (d), we plot $\Omega_-(\bm{k})$ associating with the lower magnon bands in (a) and (b). The Berry curvature exhibits sharp peaks around where the band gap $\hbar|\omega_+-\omega_-|$ reaches local minima. Integrating $\Omega_-(\bm{k})$ over the Brillouin zone gives the Chern number of the lower band, which is $0$ and $1$ for the two cases, respectively. This confirms that the system can indeed exhibit different topological phases by varying the NN exchange interactions.

To better understand the underlying physics of the two topological phases, we turn to two limits: $J_2=J_3\rightarrow0$ and $J_2=J_3\gg1$. If we turn off two of the NN exchange interactions (\textit{e.g.}, by setting $J_2=J_3=0$) as well as $D$, the system breaks up into an array of isolated diatomic pairs each bonded only along the $\alpha_1$ direction. Since neighboring pairs do not talk, the whole system has two exact flat bands separated by a large trivial gap of $\Delta=2J_1S$. 
Reintroducing $D$ is inadequate to close this gap, and in turn, change the topology. By contrast, when $J_2=J_3\gg1$, the system can be effectively viewed as multiple one dimensional SSH chains along the $\bm{\beta}_2$ direction that are weakly coupled through $J_1$. Even though individual SSH chains are gapless, the DMI will open a band gap and give rise to a nontrivial band topology.

\begin{figure}[t]
  \centering
  \includegraphics[width=\linewidth]{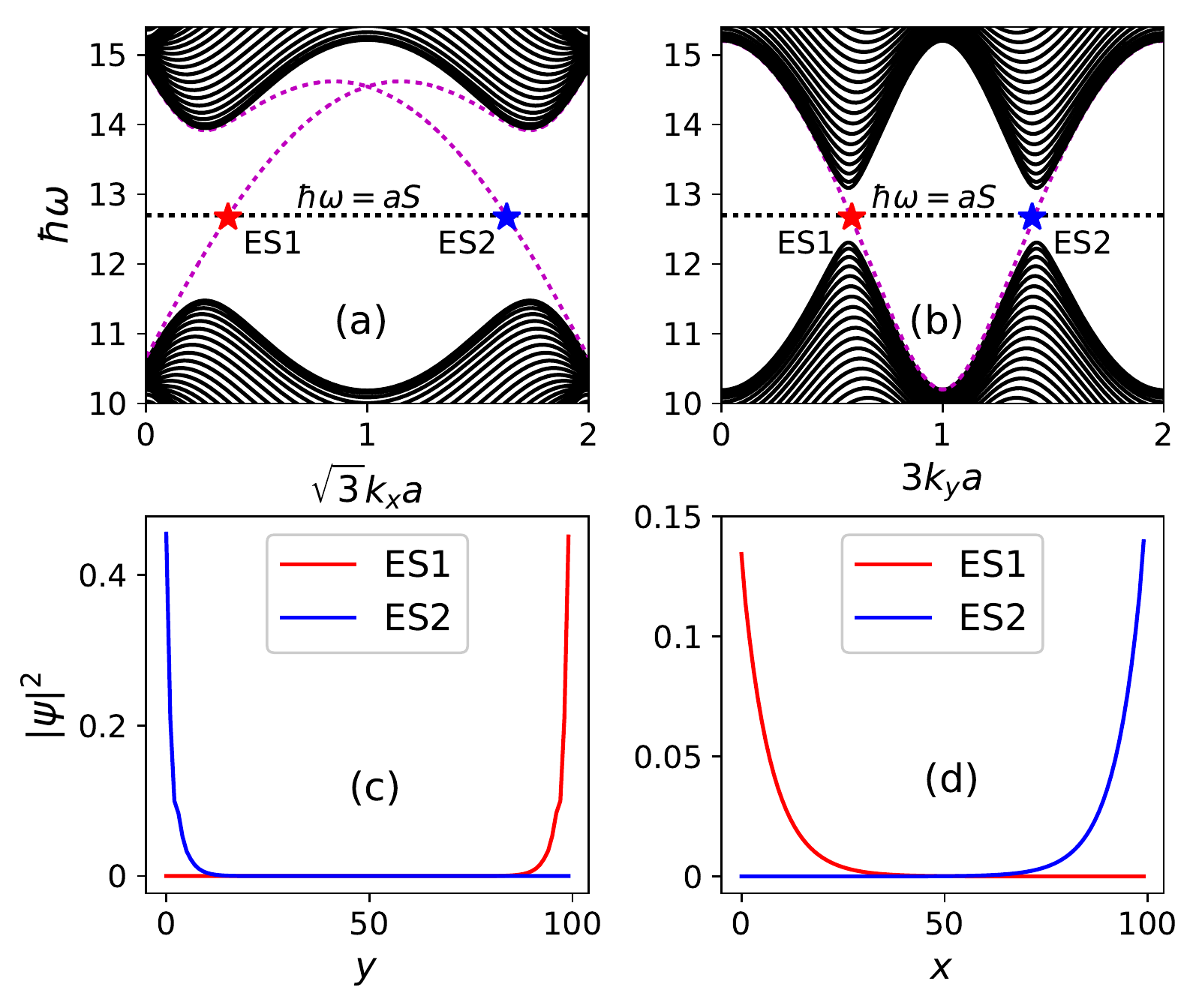}
  \caption{(color online).
  Band structure with a pair of chiral edge modes lying in the bulk gap for the armchair edge (a) and the zigzag edge (b). (c) and (d) are the spatial profile (magnon density) of the chiral edge states at the intersection with the $\hbar\omega=a$ line, labeled as ES1 and ES2. The width of the ribbon is taken as $W=100$, and $J_2=J_3=2$.
  	}
\label{edge}
\end{figure}

\textit{Edge states.---}
Guaranteed by the universal bulk-edge correspondence, a non-trivial band topology is always accompanied by chiral edge states~\cite{Liu2016The}. The phase characterized by $C=1$ [Fig.~\ref{bc} (b) and (d)] corresponds to a MHI, which supports chiral edge modes on open boundaries. To better visualize the edge states, we now wrap up the honeycomb sheet into a ribbon so that the system is periodic in one direction, leaving open boundaries in the other. Based on the crystal structure in Fig.~\ref{schematic}(a), the ribbon has a zigzag boundary along $x$ while an armchair boundary along $y$~\cite{Supply}. Figure.~\ref{edge} (a) and (b) show the topologically-protected edge modes associated with each type of open boundaries. Although the bulk band dispersion in the zigzag direction is manifestly different from that in the armchair direction, the edge modes always connect the lower and the upper bulk bands, which cannot be broken perturbatively. Moreover, regardless of the type of boundaries, the edge modes always appear in pairs and intersect the $\hbar\omega=a$ line with opposite slops $v_g=\partial{\omega}/\partial{k}$ (marked by red and blue asterisks). In other words, they propagate in opposite directions.

To see why the edge states are chiral, we plot the wavefunctions of the edge states in Fig.~\ref{edge} (c) and (d). In both types of open boundaries, it turns out that edge states of opposite group velocity are localized on opposite sides. Consequently, the propagation direction of edge magnons is locked to the side of the open boundaries. The spatial extension of an edge state can roughly be evaluated as $w\sim\left|v_g\right|/\delta$ with $\delta$ being the bulk band gap~\cite{Konig2008The}, which explains why the chiral edge states on the zigzag boundary [Fig.~\ref{bc} (b) and (d)] penetrate deeper into the bulk compared to those of the armchair boundary [Fig.~\ref{bc} (a) and (c)]. In sharp contrast to the $C=1$ phase, the $C=0$ phase is not accompanied by any topologically-protected edge states as shown in the SI~\cite{Supply}.

\begin{figure}[t]
  \centering
  \includegraphics[width=\linewidth]{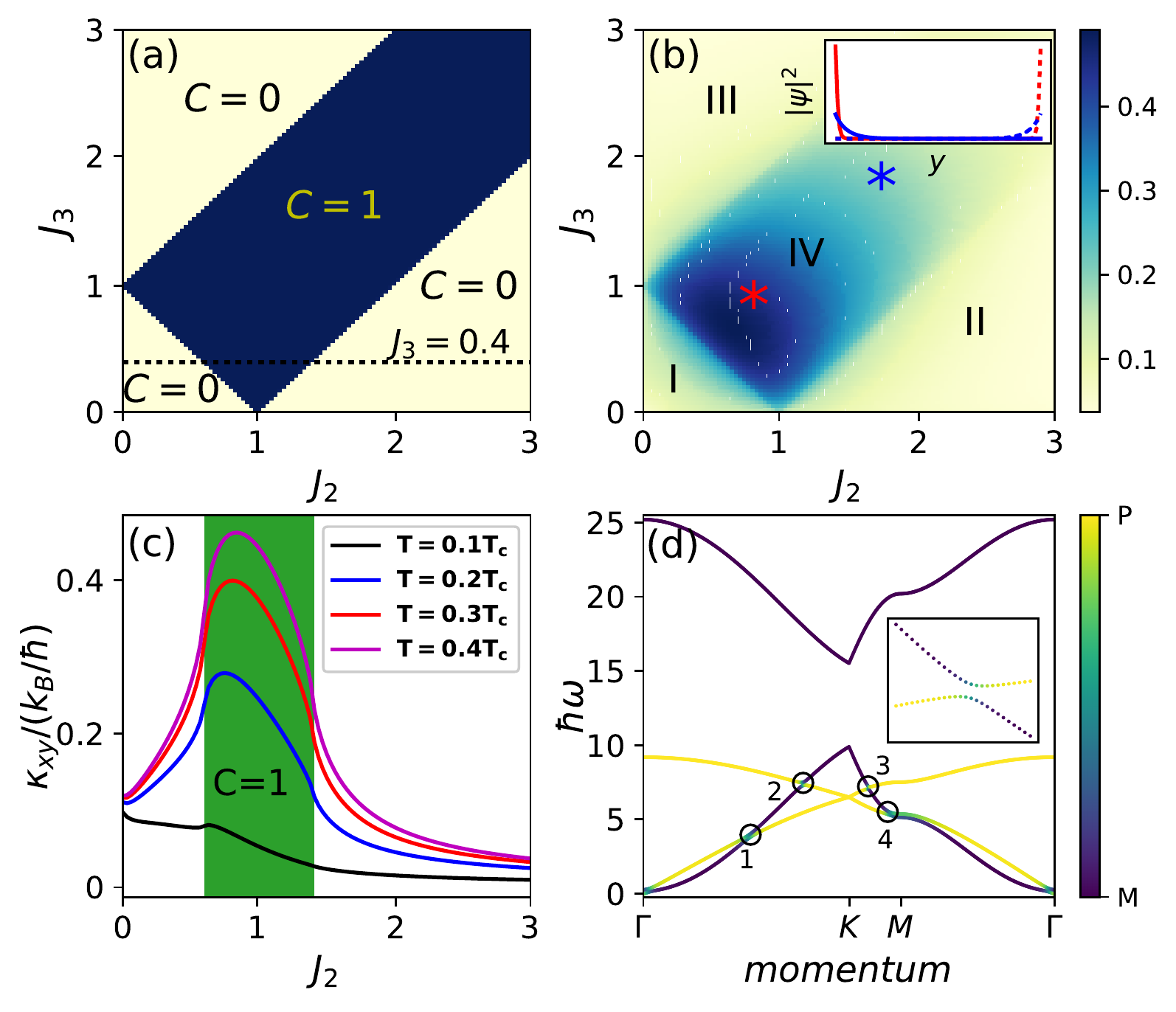}
  \caption{(color online).
  	(a) Chern number of the lower magnon band on the $J_2-J_3$ plane for $D=0.1$; the phase boundary is independent of $D$.
	(b) Thermal Hall coefficient $\kappa_{xy}$ as a function of $J_{2}$ and $J_{3}$ with temperature $T=0.4T_c$. The inset shows the edge states with energy $\hbar\omega=a$ for the two asterisked points in region IV.
	(c) $\kappa_{xy}$ as a function of $J_{2}$ at different temperatures for $J_{3}=0.4$, corresponding to the dotted line in (a).
	(d) Renormalized band dispersion along the $\Gamma$-$K$-$M$-$\Gamma$ loop [see Fig.~\ref{schematic} (c)] by taking into account the magnon-phonon coupling for $g_{A\pm}=g_{B\pm}=0.08$ and $C/m=1.5S$. Magnons and phonons hybridize and open tiny gaps near avoided crossing. The insert is a zoom-in of gap 3.
  	}
\label{php}
\end{figure}

\textit{Phase diagram and transport properties.---}
Having demonstrated two topologically distinct phases for different NN exchange interactions, it is nature to ask where is the phase boundary and, more importantly, what is the full phase diagram by varying $J_2$ and $J_3$ arbitrarily? In Fig.~\ref{php}, we plot the Chern number of the lower magnon band on the $J_2-J_3$ plane. The system turns out to be a MHI ($C=1$) when $J_2+J_3>1$ and $|J_2-J_3|<1$ and a trivial magnon insulator ($C=0$) otherwise. The phase boundaries are independent of the DMI so long as $D$ is nonzero~\cite{Supply}.

Even though the magnon Chern number is not directly related to quantized transport because of the bosonic statistics, different topological phases and phase transitions can still be characterized by the magnon Hall effect~\cite{Kim2016Realization,Owerre2016Topological}. In this regard, we now calculate the magnon Hall current driven by an in-plane temperature gradient, which originates from the non-zero Berry curvatures of both bands~\cite{Matsumoto2011Theoretical,Matsumoto2011Rotational,Xiao2006Berry}. Expressed in unit of a number current density, the magnon Hall current is
 \begin{align}
\bm{j}^H&=\frac{k_B}{\hbar}\hat{\bm{z}}\times\bm{\nabla}{T}\sum_{n=\pm}\int\frac{d^2k}{(2\pi)^2}\Omega_n\left(\bm{k}\right)\notag\\
&\left\{\rho_n\left(\bm{k}\right)\text{ln}\rho_n\left(\bm{k}\right)-\left[1+\rho_n\left(\bm{k}\right)\right]\text{ln}\left[1+\rho_n\left(\bm{k}\right)\right]\right\},
\end{align}
where $k_{B}$ is the Boltzmann constant and $\rho_n\left(\bm{k}\right)=1/\left(e^{\hbar\omega_n(\bm{k})/k_{B}T}-1\right)$ is the Bose-Einstein distribution function. Since magnons carry both spins and heat, the magnon Hall effect involves simultaneously a spin Hall current a thermal Hall current. 

Fig.~\ref{php}(b) plots the magnon Hall coefficient $\kappa_{xy}\equiv \partial j^H_y/\partial_x T$ on the $J_2-J_3$ plane. Besides showing the same phase boundaries as Fig.~\ref{php}(a), the phase diagram for $\kappa_{xy}$ has several unique features. First, even though $C=0$ in region I, $\kappa_{xy}$ does not vanish, which ascribes to the bosonic statistics that weights the Berry curvature non-uniformly in the Brillouin zone. Second, the system is a MHI in region IV, but $\kappa_{xy}$ decreases with an increasing $J_2+J_3$. To understand this character, we plot the wavefunctions of the edge states at two different points in region IV (marked by asterisks) in the inset, from which we can tell that the edge states from opposite boundaries strongly overlap at higher $J_2+J_3$. As a result, the backscattering of edge magnons is significantly enhanced, thus diminishing the magnon Hall effect. Furthermore, as shown in Fig.~\ref{php}(c) for $J_3=0.4$ [corresponds to the dotted line in Fig.~\ref{php}(a)], $\kappa_{xy}$ undergoes a sharp change across the phase boundaries, which becomes more prominent at higher temperatures due to enhanced magnon density. Such a mutation of the magnon Hall effect can be attributed to the emergence of chiral edge states when the system enters the MHI phase. Those distinctive features provides an unambiguous way to identify the topological phase transition experimentally.

\textit{Magnon-phonon coupling.---}
Phonons, the quanta of lattice vibrations, may also contribute to the thermal Hall effect at finite temperatures~\cite{Zhang2010Topological}. Since phonons do not carry spin angular momenta, their contribution can in principle be separated from magnons by spin-resolved measurements. However, due to the magnon-phonon interactions, the two types of quasiparticles can hybridize near band crossing, where they can no longer be individually defined~\cite{Thingstad2019Chiral,Go2019Topological,Zhang2019Thermal}. Therefore, it is imperative to check if magnon-phonon coupling substantially modify the band topology, hence the magnon Hall effect. In two dimensional materials, owing to the absence of the third dimension, out-of-plane lattice vibrations are much more amenable to thermal agitations than in-plane vibrations. Therefore, the dominant contribution stems from out-of-plane phonon modes, which greatly simplifies the problem. Using the augmented basis of $\Psi\left(k\right)=\left[\begin{matrix}a_{k},&b_{k},&c_{k-},&c_{k+}\end{matrix}\right]^{T}$ with $c_{k\pm}$ being the phonon annihilation operator, we can express the magnon-phonon coupling Hamiltonian as $\mathcal{H}_{mp}=\sum_{k}\Psi\left(k\right)^{\dagger}H_{mp}\left(k\right)\Psi\left(k\right)$, where the matrix $H_{mp}\left(k\right)$ reads~\cite{Thingstad2019Chiral}
\begin{align}
H_{mp}\left(k\right)=\left[\begin{matrix}a+d&b-ic&g_{A-}&g_{A+}\\b+ic&a-d&g_{B-}&g_{B+}\\g_{A-}^{*}&g_{B-}^{*}&\omega_{k-}^{p}\left(k\right)&0\\g_{A+}^{*}&g_{B+}^{*}&0&\omega_{k+}^{p}\left(k\right)\end{matrix}\right].
\label{mph}
\end{align}
Here, $g_{\alpha\pm}$ is the (phenomenological) coupling strength between the $\alpha(=A,B)$ sublattice and the $\pm$ phonon mode, and $\omega_{k\pm}^{p}\left(k\right)=\sqrt{\frac{C}{m}}\sqrt{3\pm\sqrt{3+2S_{k}}}$ is the free phonon dispersion with $C$ the elastic constant, $m$ the effective mass of the lattice site and $S_{k}=\sum_{i=1}^{3}{\text{cos}}\left(\bm{k}\cdot\bm{\beta}_{i}\right)$.

Using typical magnon-phonon coupling strengths, we diagonalize the total Hamiltonian including Eq.~\eqref{mph} and plot the band structure in Fig.~\ref{php} (d), where the magnon and phonon characters of the eigenmodes are indicated by different colors. The magnon-phonon hybridization results in avoided crossings in the lower magnon band. In the vicinity of an avoided crossing, the Berry curvature transfers from one band to the other~\cite{Thingstad2019Chiral}. Since phonons do not carry spins, the phonon contribution to the magnon spin Hall current relies solely on the loss of Berry curvature in the magnon band $\hbar\omega_-$ near these avoided crossings. Comparing Fig.~\ref{bc}(d) with Fig.~\ref{php}(d), however, we can tell that none of the avoided crossings takes place near the concentration of the Berry curvature unless the magnon-phonon coupling becomes unreasonably large such that the anti-crossings span a wide range of momenta in the Brillouin zone. Consequently, phonons only causes negligibly small modifications to the magnon Hall current; the transverse spin current is thus robust against weak magnon-phonon coupling.

\textit{Candidate materials.---}Transition metal trichalcogenides such as CrI$_3$~\cite{Huang2017Layer} and Cr$_2$Ge$_2$Te$_6$~\cite{Gong2017Discovery} exhibit stable ferromagnetism down to the monolayer limit. In these materials, magnetic atoms are arranged on a honeycomb lattice, pointing perpendicular to the plane. Because the NN exchange interactions depend sensitively on the inter-atomic distances~\cite{Ulloa2017Piezospintronic}, applying mechanical strains to the lattice~\cite{Webster2018Strain,Wu2019Strain,Aitken2010Effects} can be a viable way to induce appreciable changes to the NN exchange interactions. Under lattice deformation, the second-NN DMI may also acquire directional anisotropy as illustrated in Fig.~\ref{schematic}(b). Nevertheless, as long as the DMI does not change sign, the topology of magnon bands remains the same, which is detailed in the SI~\cite{Supply}.

In summary, we have theoretically demonstrated a magnonic analog of the SSH model and obtained the topological phase diagram of magnons characterized by both the Chern number of the lower band and the magnon Hall coefficient. The chiral edge modes result in sharp changes of the magnon Hall effect across the phase boundaries, providing a smoking-gun signature for experimental detection. Considering weak magnon-phonon coupling does not change the essential conclusion.

This work is supported by the startup fund of the University of California, Riverside.

\bibliography{ref}

\end{document}


\title{Supplemental material for ``Magnetic Su-Schriffer-Heeger model in honeycomb ferromagnets"}

\author{Yu-Hang Li}
\affiliation{Department of Electrical and Computer Engineering, University of California, Riverside, California 92521, USA}
\author{Ran Cheng}
\email[]{rancheng@ucr.edu}
\affiliation{Department of Electrical and Computer Engineering, University of California, Riverside, California 92521, USA}
\affiliation{Department of Physics, University of California, Riverside, California 92521, USA}

\date{\today}
\maketitle

\section{I. Edge states.}
\begin{figure}[htbp]
  \centering
  \includegraphics[width=0.8\textwidth]{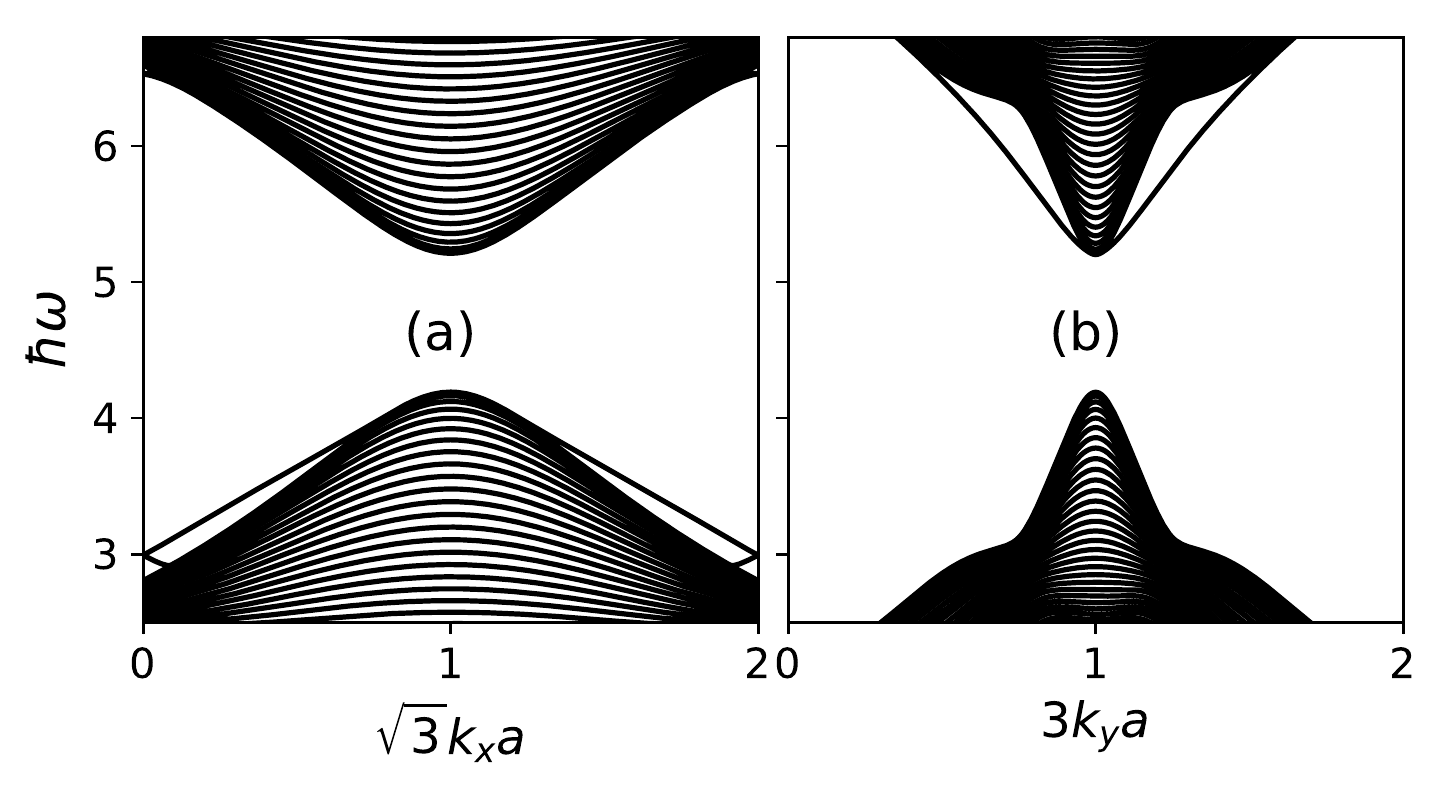}
  \caption{(color online).
             (a) and (b) are band dispersions of one dimensional nanoribbon with armchair and zigzag boundaries, respectively. Here, the width of the ribbon is $W=100$, and $J_2=J_3=0.4J_1$.
             }
\label{edge}
\end{figure}

The edge states can be studied in a nanoribbon structure {\cite{Neto2009}}. To this end, we first consider a zigzag boundary and rewrite the Eq.(1) in the main text in terms of the lattice indices $m$ and $n$,
\begin{align}
\begin{split}
\mathcal{H}_z=&\sum_{mn}\left\{-J_1\bm{S}_A\left(m,n\right)\bm{S}_B\left(m,n+1\right)-J_2\bm{S}_A\left(m,n\right)\bm{S}_B\left(m,n\right)-J_3\bm{S}_A\left(m,n\right)\bm{S}_B\left(m-1,n\right)\right.\\
&\quad+D\hat{z}\cdot\left[\bm{S}_A\left(m,n\right)\times\bm{S}_A\left(m,n+1\right)-\bm{S}_A\left(m,n\right)\times\bm{S}_A\left(m+1,n+1\right)\right.\\
&\qquad\quad+\bm{S}_A\left(m,n\right)\times\bm{S}_A\left(m+1,n\right)-\bm{S}_B\left(m,n\right)\times\bm{S}_B\left(m,n+1\right)\\
&\qquad\quad\left.+\bm{S}_B\left(m,n\right)\times\bm{S}_B\left(m+1,n+1\right)-\bm{S}_B\left(m,n\right)\times\bm{S}_B\left(m+1,n\right)\right]\\
&\quad+\kappa\left[S_A^z\left(m,n\right)^2+S_B^z\left(m,n\right)^2\right]\left.\right\},
\label{eq1}
\end{split}
\end{align}
where, $J_{i=1,2,3}$, $D$ and $\kappa$ share the same meanings as those in the main text. Using the Holstein-Primakoff transformation
\begin{align}
\begin{split}
&S_A^+\left(m,n\right)=\sqrt{2S}a_{mn},\qquad S_A^-\left(m,n\right)=\sqrt{2S}a^\dagger_{mn}, \qquad S_A^z\left(m,n\right)=S-a_{mn}^\dagger a_{mn},\\
&S_B^+\left(m,n\right)=\sqrt{2S}b_{mn},\qquad S_B^-\left(m,n\right)=\sqrt{2S}b^\dagger_{mn}, \qquad S_B^z\left(m,n\right)=S-b_{mn}^\dagger b_{mn},
\end{split}
\end{align}
the Hamiltonian in Eq.~\eqref{eq1} can be written as
\begin{align}
\begin{split}
\mathcal{H}_{z}/S=&-J_1\sum_{m=1}^W\left(a_{mn}b_{mn+1}^\dagger+a_{mn}^\dagger b_{mn+1}-a_{mn}^\dagger a_{mn}-b_{mn}^\dagger b_{mn}\right)\\
&-J_2\sum_{m=1}^W\left(a_{mn}b_{mn}^\dagger+a_{mn}^\dagger b_{mn}-a_{mn}^\dagger a_{mn}-b_{mn}^\dagger b_{mn}\right)\\
&-J_3\sum_{m=2}^W\left(a_{mn}b_{m-1n}^\dagger+a_{mn}^\dagger b_{m-1n}-a_{mn}^\dagger a_{mn}-b_{m-1n}^\dagger b_{m-1n}\right)\\
&+\frac{D}{2i}\left[\sum_{m=1}^W\left(a_{mn}^\dagger a_{mn+1}-a_{mn}a_{mn+1}^\dagger-b_{mn}^\dagger b_{mn+1}-b_{mn}b_{mn+1}^\dagger\right)\right.\\
&\quad\ \ -\sum_{m=1}^{W-1}\left(a_{mn}^\dagger a_{m+1n+1}-a_{mn}a_{m+1n+1}^\dagger-b_{mn}^\dagger b_{m+1n+1}-b_{mn}b_{m+1n+1}^\dagger \right)\\
&\quad\ \ +\sum_{m=1}^{W}\left(a_{mn}^\dagger a_{m+1n}-a_{mn}a_{m+1n}^\dagger -b_{mn}^\dagger b_{m+1n}-b_{mn}b_{m+1n}^\dagger \right)\left.\right]\\
&+2\kappa\sum_{m=1}^{W-1}\left(a_{mn}^\dagger a_{mn}+b_{mn}^\dagger b_{mn}\right),
\end{split}
\end{align}
where the summation of index $n$ has been omitted for simplification. Since we use a periodic boundary condition in the $x$ direction, after the Fourier transformation, $\psi_{kn}\equiv\left(\begin{matrix}a_{mn}\\b_{mn}\end{matrix}\right)=1/\sqrt{N}\sum_{k}e^{ikn}\left(\begin{matrix}a_{mk}\\b_{mk}\end{matrix}\right)$, the Hamiltonian can finally be written as
\begin{align}
\begin{split}
\mathcal{H}_{z}/S=&\sum_{m=1}^{W}\left[\left(J_{1}+J_{2}-2\kappa\right)\left(a_{mk}^\dagger a_{mk}+b_{mk}^\dagger b_{mk}\right)-\left(J_{1}e^{\sqrt{3}ika}+J_{2}\right)\left(a_{mk}b_{mk}^\dagger +\text{H.c.}\right)\right]\\
&-2D\sin{\sqrt{3}ka}\sum_{m=1}^{W}\left(a_{mk}^\dagger a_{mk}-b_{mk}^\dagger b_{mk}\right)+J_{3}\sum_{m=2}^{W}\left[\left(a_{mk}^\dagger a_{mk}+b_{mk}^\dagger b_{mk}\right)-\left(a_{mk}b_{m-1k}^\dagger +\text{H.c.}\right)\right]\\
&+iD\sum_{m=2}^{W}\left[\left(e^{-\sqrt{3}ika}-1\right)\left(a_{mk}^\dagger a_{m+1k}-b_{mk}^\dagger b_{m+1k}\right)-\text{H.c.}\right],
\label{zigzag}
\end{split}
\end{align}
where the summation of $k$ is omitted. By the same token, the Hamiltonian of a nanoribbon with an armchair boundary can finally be written as
\begin{align}
\begin{split}
\mathcal{H}_{a}/S=&\sum_{m=1}^{W}\left[\left(J_{3}-2\kappa\right)\left(a_{mk}^\dagger a_{mk}+b_{mk}^\dagger b_{mk}\right)-2D\sin{3ka}\left(a_{mk}^\dagger a_{mk}-b_{mk}^\dagger b_{mk}\right)-J_{3}\left(a_{mk}b_{mk}^\dagger +\text{H.c.}\right)\right]\\
&+J_{1}\sum_{m=1}^{W-1}\left[\left(a_{mk}b_{m+1k}^\dagger e^{3ika}+\text{H.c.}\right)-\left(a_{mk}^\dagger a_{mk}+b_{mk}^\dagger b_{mk}\right)\right]\\
&+J_{2}\sum_{m=2}^{W}\left[\left(a_{mk}b_{m-1k}^\dagger +\text{H.c.}\right)-\left(a_{mk}^\dagger a_{mk}+b_{mk}^\dagger b_{mk}\right)\right]\\
&+iD\sum_{m=1}^{W-1}\left[\left(e^{-3ika}+1\right)\left(a_{mk}^\dagger a_{m+1k}-b_{mk}^\dagger b_{m+1k}\right)-\text{H.c.}\right].
\label{armchair}
\end{split}
\end{align}
Diagonalizing Eqs.~\eqref{zigzag} and \eqref{armchair}, the energy dispersions and the corresponding eigenvectors can be derived directly, base on which the density distribution functions in the real space can be obtained straightforwardly. Figs.~\ref{edge} (a) and (b) show the one dimensional band dispersions of a normal magnon insulator in a nanoribbon with armchair and zigzag boundaries, respectively, where the bands are well gaped without any survived edge states.

\section{II. Different Dzyaloshinskii-Moriya interactions.}
In the main text, we propose that the topological phase transition can be tested in honeycomb ferromagnets under proper deformations. Nevertheless, such deformations generally change not only the nearest-neighbor exchange interactions but also the next nearest-neighbor Dzyaloshinskii-Moriya interactions. In this case, the minimal Hamiltonian reads
\begin{equation}
\mathcal{H}=-\sum_{\left<i,j\right>}J_{ij}\bm{S}_{i}\cdot\bm{S}_{j}+\sum_{\left<\left<i,j\right>\right>}D_{ij}\epsilon_{ij}\hat{z}\cdot\bm{S}_{i}\times\bm{S}_{j}+\kappa\sum_{i}S_{iz}^{2},
\end{equation}
where the Dzyaloshinskii-Moriya interactions take $D_{1}=D\cdot J_{1}J_{2}$ along $\beta_{1}$, $D_{2}=D\cdot J_{2}J_{3}$ along $\beta_{2}$, and $D_{3}=D\cdot J_{1}J_{3}$ along $\beta_{3}$, respectively. All other parameters share the same meanings and take the same values as those in the main text. 

\begin{figure}[htbp]
  \centering
  \includegraphics[width=0.9\textwidth]{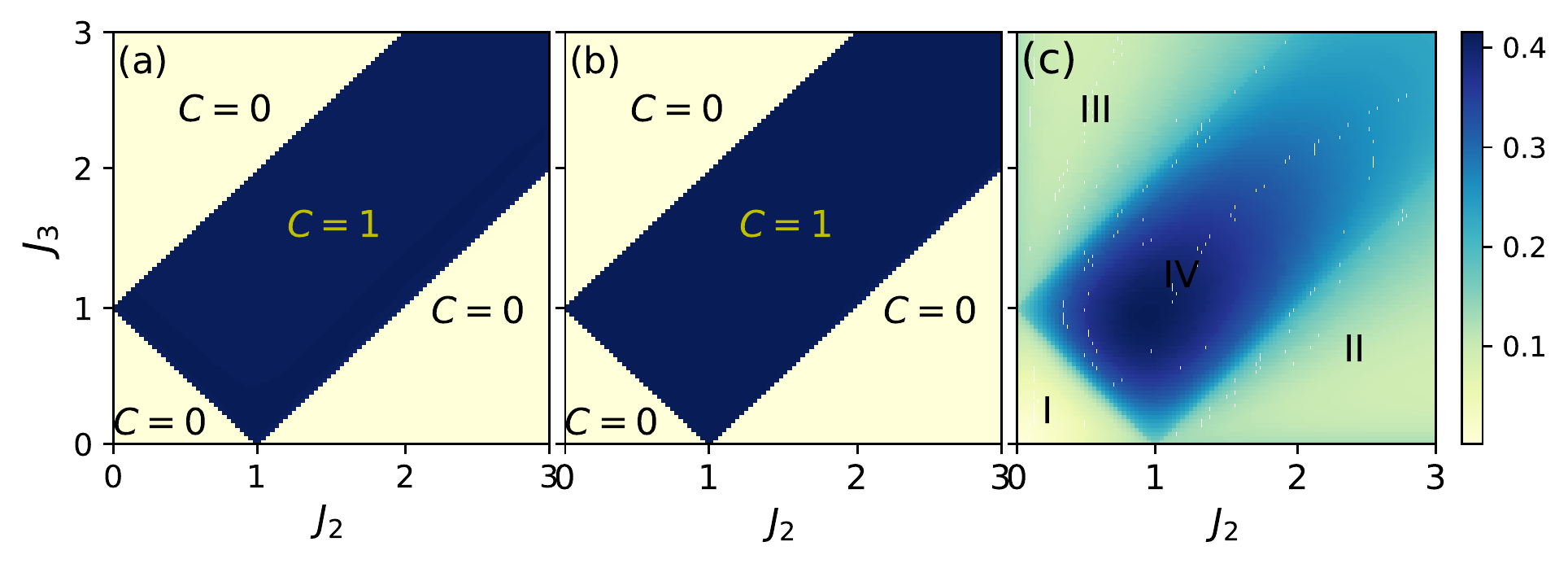}
  \caption{(color online).
             (a) Chern number of the lower magnon band on the $J_2-J_3$ plane with $D_1=D_2=D_3=0.2$.
             (b) Chern number of the lower magnon band on the $J_2-J_3$ plane when taking different Dzyaloshinskii-Moriya interactions into consideration.
             (c) Thermal Hall coefficient $\kappa_{xy}$ corresponding to (b) on the $J_2-J_3$ plane with temperature $k_BT = 0.5J_1$ . All other parameters are exactly the same as those in the main text. 
             }
\label{dmi}
\end{figure}

As a comparison, Fig.~\ref{dmi} (a) plots the Chern number of the lower magnon band on the $J_2-J_3$ plane with $D_1=D_2=D_3=0.2$ while Fig.~\ref{dmi} (b) shows the same band Chern number when taking different DMI into consideration.  Both phase diagrams are exactly the same with that in the main text, confirming that the total magnon band topology does not depend on the detailed values of different Dzyaloshinskii-Moriya interactions unless they are nonzero. Here is the reason. The nonzero Dzyaloshinskii-Moriya interactions induce an inverted band gap between the lower and upper bands, which results in a topological nontrivial band {\cite{Bansil2016}}. The band topology should not change with the difference between the Dzyaloshinskii-Moriya interactions, because this inverted band gap can not be closed by the Dzyaloshinskii-Moriya interactions as long as they do not change sign. In addition to that, the specific band dispersions may be modified by the Dzyaloshinskii-Moriya interactions, and consequently, the thermal Hall coefficient $\kappa_{xy}$ as shown in Fig.~\ref{dmi} (c) may be quantitatively different now.
However, the sharp increase across the phase boundary from normal magnon insulators to the magnon Hall insulator remains an obvious signature. Therefore, the thermal Hall effect is still an efficient method to detect this topological phase transition in experiment.

